\documentclass[pdflatex,a4paper]{article}

\usepackage{float}
\newfloat{SupplementaryTable}{h!}{ext}
\floatname{SupplementaryTable}{Supplementary Table}

\usepackage{amsmath}
\usepackage{amssymb}

\usepackage{bibunits}
\defaultbibliographystyle{unsrt}

\usepackage{times}
\usepackage{graphicx}
\usepackage{geometry}

\usepackage{authblk}
\usepackage{hyperref}

\begin{document}

\title{On the Transmission of Texts:\\ Written Cultures as Complex Systems}

\author[1]{Jean-Baptiste Camps} 
\author[2,3]{Julien Randon-Furling}
\author[1]{Ulysse Godreau}

\affil[1]{École nationale des chartes, Paris, Sciences \& Lettres, 65 rue de Richelieu, Paris, France}

\affil[2]{Centre Borelli, ENS Paris-Saclay, Univ. Paris-Saclay, Paris, France}

\affil[3]{College of Computing UM6P, Ben Guerir, Morocco}

\date{}

\maketitle

\begin{abstract}
Our knowledge of past cultures relies considerably on written material. For centuries, texts have been copied, altered, then transmitted or lost --~eventually, from surviving documents, philologists attempt to reconstruct text phylogenies (``\textit{stemmata}’’), and past written cultures. Nonetheless, fundamental questions on the extent of losses, representativeness of surviving artefacts, and the dynamics of text genealogies have remained open since the earliest days of philology.\\
To address these, we radically rethink the study of text transmission through a complexity science approach, integrating stochastic modelling, computer simulations, and data analysis, in a parsimonious mindset akin to statistical physics and evolutionary biology. Thus, we design models that are simple and general, while accounting for diachrony and other key aspects of the dynamical process underlying text phylogenies, such as the extinction of entire branches or trees. On the well-known case study of Medieval French chivalric literature, we find that up to 60\% of texts and 99\% of manuscripts were lost (consistent with recent synchronic ``biodiversity'' analyses). We also settle a hundred-year-old controversy on the bifidity of stemmata. Further, our null model suggests that pure chance (``\textit{drift}'') is not the only mechanism at play, and we provide a theoretical and empirical framework for future investigation.
\end{abstract}

\maketitle

\begin{bibunit}
\section{Introduction}

How did written works survive~\cite{cisne_how_2005}?  Why do we know the names of Gilgamesh, Ulysses or King Arthur, and how many others have we forgotten?
How much do we preserve of the written knowledge and cultures of the past? And how representative is what we know, compared to what existed? Such fundamental questions depend on the actual and material process through which texts were produced, distributed and ultimately transmitted or lost. 

Before the advent of the printing press, written texts were circulated in manuscript form. 
In order to make the text available, the author would dictate it to a secretary, or write a draft on wax tablets, papyrus, parchment or, later, paper, and this original, authorial, manuscript would then have to be copied manually by a scribe in the form of a new manuscript, and then circulated. Copies could in turn be used to create more manuscripts, again by manual copying, perhaps by other scribes in other regions at a later date. During this process, successive innovations were introduced in the text, either mistakenly, or intentionally to make the text more suited to its intended audience.  
These alterations in the written sequence forming the text could then be transmitted to its ``descendants'' by a given manuscript. 
Wear and tear, accidents, and fashions caused the destruction of some manuscripts, while others enjoyed the long life of library preservation. In the end, knowledge was lost, textual diversity reduced, and some texts went extinct, while others gained traction and were eventually preserved for future generations.

Scholars have tried to make sense of the surviving documents, by ascertaining the network of relationships between them, by making assumptions about lost sources or by trying to reconstruct such lost sources. In the process of doing so, they have produced evolutionary trees that philologists call \textit{stemmata}. Yet, these stemmata only reflect the portion of the trees that are accessible to us by surviving evidence, while loss of documents can prune away entire branches (Figure~\ref{fig:fig1}, \textbf{A} and \textbf{B}). 
For centuries, historians and philologists working on the transmission of medieval texts have found their efforts counteracted by a gaping hole in existing documentation ---~a hole the actual size of which remained unknown: how much of what actually existed do surviving works and documents represent? 

A variety of methods have been used to estimate the loss of documents, drawing from the items surveyed in medieval catalogues \cite{buringh_medieval_2010}, or using methods inspired by paleodemography~\cite{cisne_how_2005} or ecology. 
In particular, recent work by Kestemont et al.~\cite{kestemont_estimating_2020,kestemont_forgotten_2022} used unseen species methods to estimate the ``unseen'' (i.e. lost) part of medieval chivalric literatures, treating works as species,  individuals as documents, and contemporary libraries as observation sites. Doing so, they produced plausible estimates, that have received wide attention. Yet, the unseen species method takes the point of view of contemporary libraries and treats manuscripts that were produced at very different dates as equivalent individuals; as such, it does not take into account the dynamic nature of the process of textual transmission and evolution in time. Manuscripts were simultaneously being copied and destroyed during hundreds of years; effects and biases leading to the survival or extinction of different branches in the tree of a text are hard to predict.
In this sense, the problem of transmission, survival and extinction of texts is one of loss of evolutionary history, as is encountered also in evolutionary biology \cite{yessoufou_reconsidering_2016}: how does the loss of nodes and branches affect the trees of texts, that are known to exhibit peculiar properties such as bifidity and imbalance?\\ We present here a stochastic modelling and simulation framework that accommodates dynamical phenomena while allowing us to investigate the aforementioned properties and to estimate observed as well as unobserved variables. In so doing, we take what one may call a complexity science approach to philology. 

\section{A philological forest of imbalanced trees} \label{sec:philo_for_imbal}

\begin{figure}[phtb]
    \centering %
    \includegraphics[width=\textwidth]{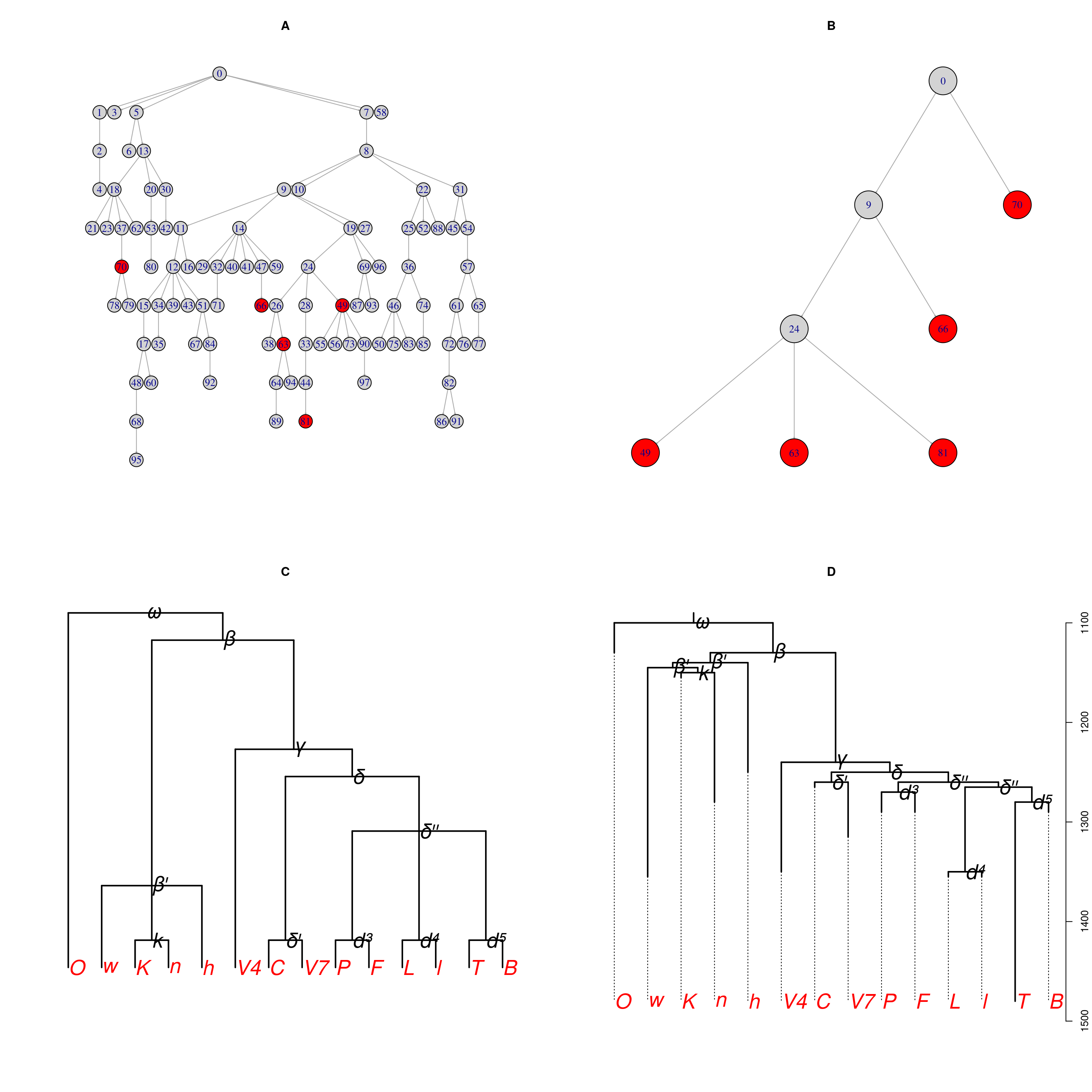}
    \caption{\label{fig:fig1} \textbf{From the real tree to the \textit{stemma codicum}}. The stemma will contain all witnesses, and as many hypothetical nodes as are needed to explain their relationships.
    \textbf{A.}~Simulated complete tree of a textual tradition, red nodes depict surviving manuscripts (i.e.~witnesses), and grey nodes lost ones; \textbf{B.}~reduced tree (\textit{stemma codicum}) showing only witnesses, and their latest shared ancestors; this corresponds to the part of the original tree that could be inferred by philologists from surviving evidence;
    \textbf{C.}~ Stemma of the \textit{Song of Roland}, following Segre~\cite{segre_chanson_1971}, roman letters (in red) stand for extant witnesses, and greek letters for hypothetical shared ancestors; stemmata contain polytomies (nodes with out-degree greater than 2) when a given ancestor has been copied more than twice, and more than two lineages of its descendants survive;
    \textbf{D.}~in practice, a stemma codicum could be turned into a binary phylogram, by introducing time, and considering successive copies of a same manuscript as distinct events.%
    }
\end{figure}

Philology studies the transmission and evolution of texts. Some of its classic models were introduced in the 19th century, contemporaneously to Darwin's work on the evolution of species and similar evolutionary models in historical linguistics~\cite{schlyter_corpus_1827}.
One of the staple methods introduced then is the so-called ``common errors'' method~\cite{timpanaro_genesi_2003,haugen_2_2020}, whereby common innovations in the text appearing in surviving copies (called \textit{witnesses}) of a work are exploited in order to infer genealogy-like relationships. From these, a tree-type graph is obtained for the \textit{tradition} (the set of manuscripts that collectively preserve a given work), and this graph is called a \textit{stemma codicum}~(Fig.~\ref{fig:fig1}, \textbf{C}). This is of course much reminiscent of the way phylogenetic trees or cladograms are obtained in biology, based on shared characteristics between observed or inferred species, and it is possible to turn a stemma into a binary phylogram~(Fig.~\ref{fig:fig1}, \textbf{D}). Methods from cladistics and phylogenetics  have sometimes been directly applied to texts, with results that remain debated~\cite{barbrook_phylogeny_1998,howe_manuscript_2001,spencer_phylogenetics_2004,mace_evolution_2006,hoenen_history_2020}. 

An interesting observation about the structure of stemmata\footnote{It is customary to use the Latin form of the plural for \textit{stemma}, hence \textit{stemmata}.} was made by French philologist Joseph Bédier, almost a century ago, in 1928~\cite{bedier_tradition_1928}: most trees reconstructed by philologists exhibit root bifurcation, i.e.~a root with out-degree~$2$. In most reconstructions, Bédier observed, the original or the last common ancestor (LCA, commonly called \textit{archetype}) has two, and only two, direct descendants. This prevents an accurate reconstruction of the original text by majority principle (e.g.~two witnesses vs one in favour of a given variant). Discarding without further ado the possibility that such dominance of degree-$2$ roots might actually emerge from the dynamics of text transmission, Bédier immediately interpreted this ``forest of bifid trees'' as resulting from a methodological flaw or some unconscious bias, spurring a century-long debate and causing a long-lasting methodological schism within textual scholarship ---~a schism that remains to be fully resolved~\cite{baker_ombre_2018,duval_tradition_2021} and that we address here.
Modern estimates for the proportion of bifid stemmata lie somewhat lower (70\% to  83\%; Supplementary Table~\ref{tab:tree_props})
than Bédier's (95.5\%), but it is the case that a strong prevalence of root bifurcation is observed among reconstructed philological trees~\cite{shepard_recent_1930,castellani_bedier_1957,haugen_silva_2015}. 

Ubiquitous features in the genealogies inferred from surviving manuscripts, like bifidity, need not stem from biases in the reconstruction method itself, but could instead emerge from the transmission process's own dynamics.
Previous modelling research has often approached the question of bifidity by using purely combinatorial approaches or by applying decimation to static trees~\cite{greg_recent_1931,maas_leitfehler_1937,castellani_bedier_1957,guidi2004sugli,Hoenen16,hoenen_how_2017}. Yet they were generally focused specifically on the property of bifidity alone, and have left aside the dynamic nature of the transmission of texts.

Abundance of degree-$2$ roots --~and, more generally, of nodes with out-degree~$2$~\cite{haugen_silva_2015}~--, is not the only salient property of stemmata. Large asymmetry (or imbalance) between the size of branches (see Fig.~\ref{fig:fig1}~\textbf{C} and \textbf{D}), a phenomenon pervasive in evolutionary biology \cite{aldous_2001}, is also observed in many stemmata.
Moreover, absence of large parts of the original tradition, when the stemma's root is known to be not the original text but a later manuscript --~meaning that the stemma represents only a localised portion of the original tree~-- or quasi-generalised absence of direct descent between surviving witnesses are also properties potentially revealing of the transmission process and the extent of losses. In some cases, the presence of lateral transmission (generally called \textit{contamination}) originating from outside the stemma can also point to the loss of complete branches from the original tree.

Are these properties artefacts or biases from the reconstruction method, as Bédier assumed, or could they in principle emerge from the sole dynamics of text transmission, and the effects of chance (drift)? 
This calls for a null model thanks to which one could measure discrepancies between manuscript traditions and their reconstructed counterparts.
The modelling approach which is needed should therefore have the capacity 
to settle questions such as B\'edier's, and provide estimates for losses of individual manuscripts and whole traditions. Meanwhile, it should remain as simple and general as possible, because, quoting British polymath Michael~Perry~Weitzman (who introduced mathematical methods and computer simulations in philology): ``General phenomena require general explanations''~\cite{weitzman_evolution_1987}. This translates into models that are, just like in statistical physics, voluntarily sparse, almost bare --~but it is precisely this feature that makes them powerfully informative: they show how relatively simple ingredients may combine to produce complex properties that often wrongly appear at first sight extremely intricate and as necessitating a host of complicated details.

\section{Complexity science for philology}

Until now,  attempts to model the dynamic process of text transmission have remained somewhat isolated. Weitzman  articles~\cite{weitzman_computer_1982,weitzman_evolution_1987} have pioneered the use of stochastic models for manuscript transmission, but for a limited number of scenarii constraining parameters to predefined values motivated by historical extrinsic factors.   We shall in fact elaborate on Weitzman's work, but within a broader, more general framework benefiting from the present-day conceptual and computational developments of complexity science, that allows us to model, simulate and analyse the process of manuscript transmission and thus overcome the limitations of previous studies.

Complexity science~\cite{jensen22} focuses on systems that typically involve large numbers of interacting entities and exhibit a phenomenon termed \textit{emergence}~\cite{anderson72}, whereby properties at the macroscopic level appear (``emerge'') from the aggregation and the interaction of multiple events at the microscopic level ---~just as the general properties of surviving texts and of their reconstructed genealogies stem from the interplay between the individual scale (manuscripts) and the collective one (whole corpora). Complexity science draws much upon statistical physics~\cite{maxwell1873,reif67,krapiv10} and has proven increasingly effective to understand both natural 
and social systems, from spin glasses~\cite{parisi2006} and starling flocks~\cite{cavagna2010}, to organisations~\cite{cohen2000}, networks~\cite{battiston2021}, and many other examples~\cite{bouchaud2011,kiel2021,parisi2023}. 

Contemporaneously, over the past two decades an unprecedented availability of data has opened new avenues of research in multidisciplinary complex systems, perhaps most notably in the so-called \textit{digital humanities}~\cite{roth2019,schwandt2020}. Philology has partially espoused this movement, particularly for data visualisation and textual analysis across large data-bases. However, just as other fields in the humanities, philology is yet to harness the potential of the comprehensive approach that complexity science offers: intertwining physics-type modelling, including stochastic and agent-based models, with state-of-the-art computational statistics and machine-learning techniques. 

It is this path that we follow here: we use stochastic processes that are analytically tractable in their simplest form, and can always be computer simulated, in order to generate tens of thousands of artificial manuscript populations governed by the copy rate and the loss rate of manuscripts. Stemmata are then constructed for these artificial populations via an algorithm that emulates a philologist's work on surviving manuscripts. This way, we produce synthetic data allowing us to analyse manuscript transmission with tools from statistical physics and machine learning. Heatmaps are obtained for various observables (\textit{eg} the bifidity rate among stemmata) across ranges of values for the copy and loss rates. They are then compared to historical data, gathered for medieval traditions on observable features, allowing inferences for features that are not historically observable (such as loss rates). 

\subsection*{\normalsize Manuscript transmission as a stochastic process}
The specific process of text transmission --~that is, the copy and circulation of manuscripts~-- varied through time and space, from on-demand production of a single copy by an individual (amateur or professional) copyist to forms of serial production in dedicated workshops. Instead of trying to account for all details, complexity science follows statistical physics in opting for a parsimonious type of modelling: the idea being that some, if not all, of the fundamental properties of a system (especially a large system --~here tens, hundreds or even thousands of manuscripts) will most likely not depend on a host of details but rather on only a couple of variables. Hence it is relevant to start with the simplest possible model and examine what properties it already exhibits. Here we abstract all details away except the fact that manuscripts underwent copy (birth) and/or destruction (death). 

This leads to a stochastic process called a \textit{birth and death} (BD) process, a special case of a Markov chain~\cite{Norris98}. In a BD process, individuals (here manuscripts) appear (``are born'') and are lost (``die'') at certain rates,  corresponding to the \textit{copy} or \textit{birth} rate ($\lambda$) and to the \textit{loss} or \textit{death} rate ($\mu$) of manuscripts. Thus, a population initially consisting of a single original manuscript evolves through time according to the following rule: at every time step $t \rightarrow t+\Delta t$, each member of the population that is present at time $t$ has probability $\lambda\,\Delta t$ to engender a new individual (a birth event) and probability $\mu\, \Delta t$ to disappear (a death event), where rates $\lambda, \mu \in \mathbb{R}_+$ are non-negative real numbers.

\subsection*{\normalsize Generating artificial manuscript populations}

In the case where the rates $\lambda$ and $\mu$ are constant over time, simple observables of the BD process such as survival probabilities or average population are analytically tractable and may be solved exactly with pencil and paper. For more intricate observables (in particular, topological properties of the resulting trees), or when the rates $\lambda$ and $\mu$ are time-dependent, it becomes tedious to solve the model analytically. However, it may always be computer simulated as an agent-based model (ABM), which is what we do. In our ABM, agents are manuscripts, starting initially with $N$ of them. A time-step corresponds to a typical duration for a manuscript to be copied. At every time-step~$t$, each extant manuscript has probability $\lambda_t$ to be copied and probability $\mu_t$ to be lost. Keeping track of the filiations between manuscript, trees (or \emph{arbres réels}~in Fourquet's words\cite{fourquet_paradoxe_1945}) are built for each original manuscript, representing the genealogy of the full manuscript tradition stemming from that original. These trees are then reduced by only keeping the genealogical relationships between surviving manuscripts (see \emph{Methods}), giving rise to the ideal stemma that a contemporary philologist would produce.

We posit that some of the characteristics of real manuscript populations may be captured by the relatively simple stochastic process described in the previous paragraphs. To probe whether this is indeed the case, we study this process by exploring its parameter space, building heatmaps obtained by varying the parameters $\lambda$ (copy rate) and $\mu$ (loss rate) through ranges corresponding to reasonable orders of magnitude; here between $10^{-3}$ and $10^{-2}$ (\textit{Methods}). The survival rates and probability distribution of extant witnesses can be computed exactly (\textit{Methods}) as a function of the parameters. Other properties were computed by simulating a number~$N=10^4$ of artificial manuscript traditions (that is, a realisation of the stochastic process described above with $10^4$ original living manuscripts at $t=0$) for each pair of values $(\lambda,\mu)$, and for each of these traditions the corresponding stemma is constructed. We can then compute the average value of a variable of interest over the~$N$ populations: this gives the $(\lambda,\mu)$-point in the heatmap.


The empirically observable variables that we examine are:  
the median final population of surviving witnesses for traditions with at least one witness; the extant lifespan of traditions (i.e. the distance in years between the oldest and youngest surviving witnesses); the bifidity ratio of stemmata; all of which can be compared to observed data from historical traditions. Additionally, we examine 
the survival rate of traditions (i.e., trees); the survival
rate of manuscripts (i.e., nodes); as well as the median distance between the lowest common ancestor of the surviving witnesses and the actual root of the original tree (i.e. the original manuscript); all of which are not directly observable in historical data.

Subsequently, comparisons between the values of some of the same variables obtained through estimations based upon  different methods point toward a particular region in the parameter space where our artificial populations may be seen to coincide reasonably well with real manuscript populations --~as far as the estimated observables are concerned (\textit{Methods}). This allows us to use this region in the parameter space to suggest (within confidence intervals) values for variables for which we do not have estimates through other means.

\section{The Lost manuscripts and extinct texts of Charlemagne and King Arthur}

\begin{figure}
    \centering
    \includegraphics[width=0.45\textwidth]{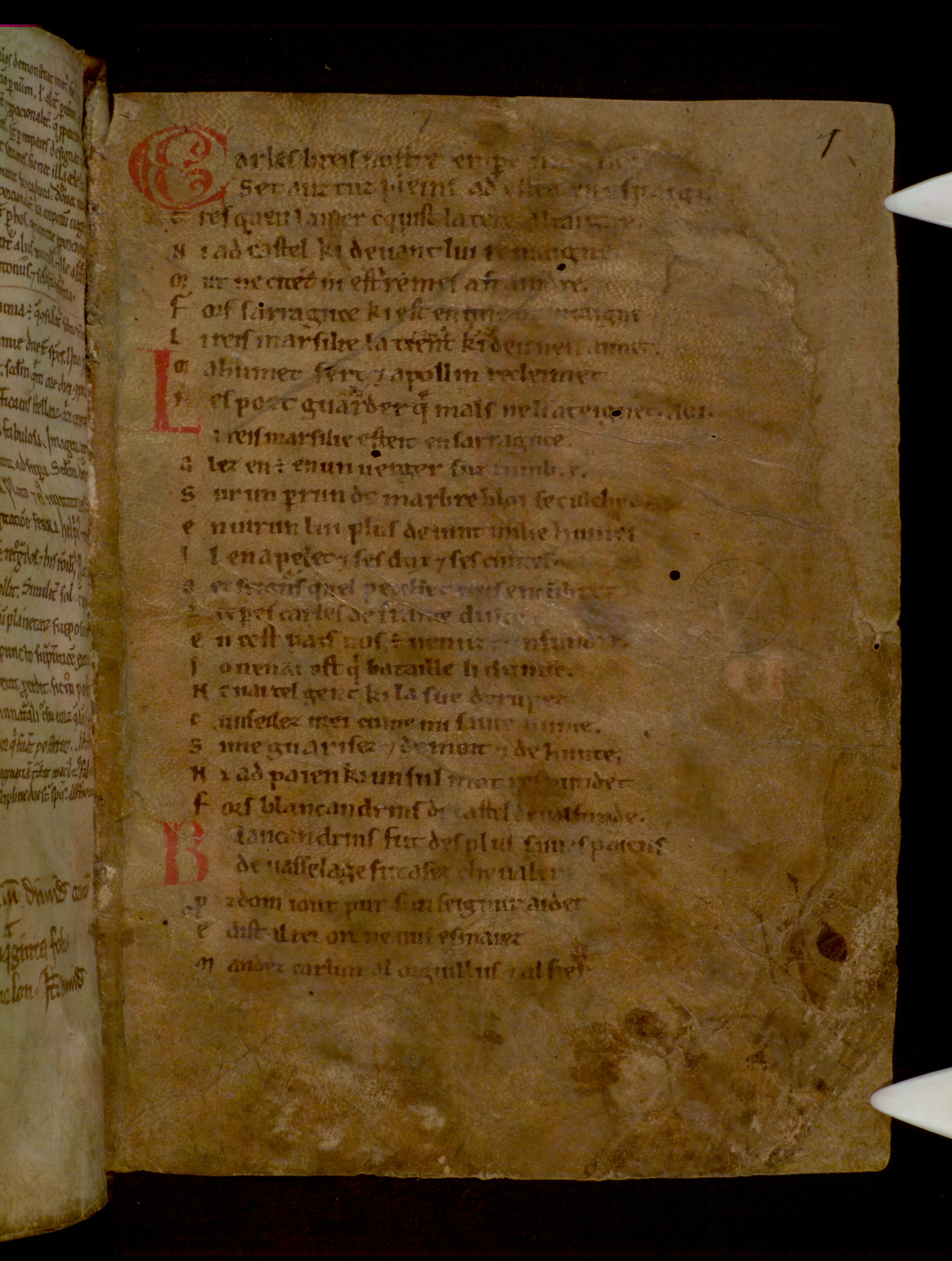}
    \includegraphics[width=0.43\textwidth]{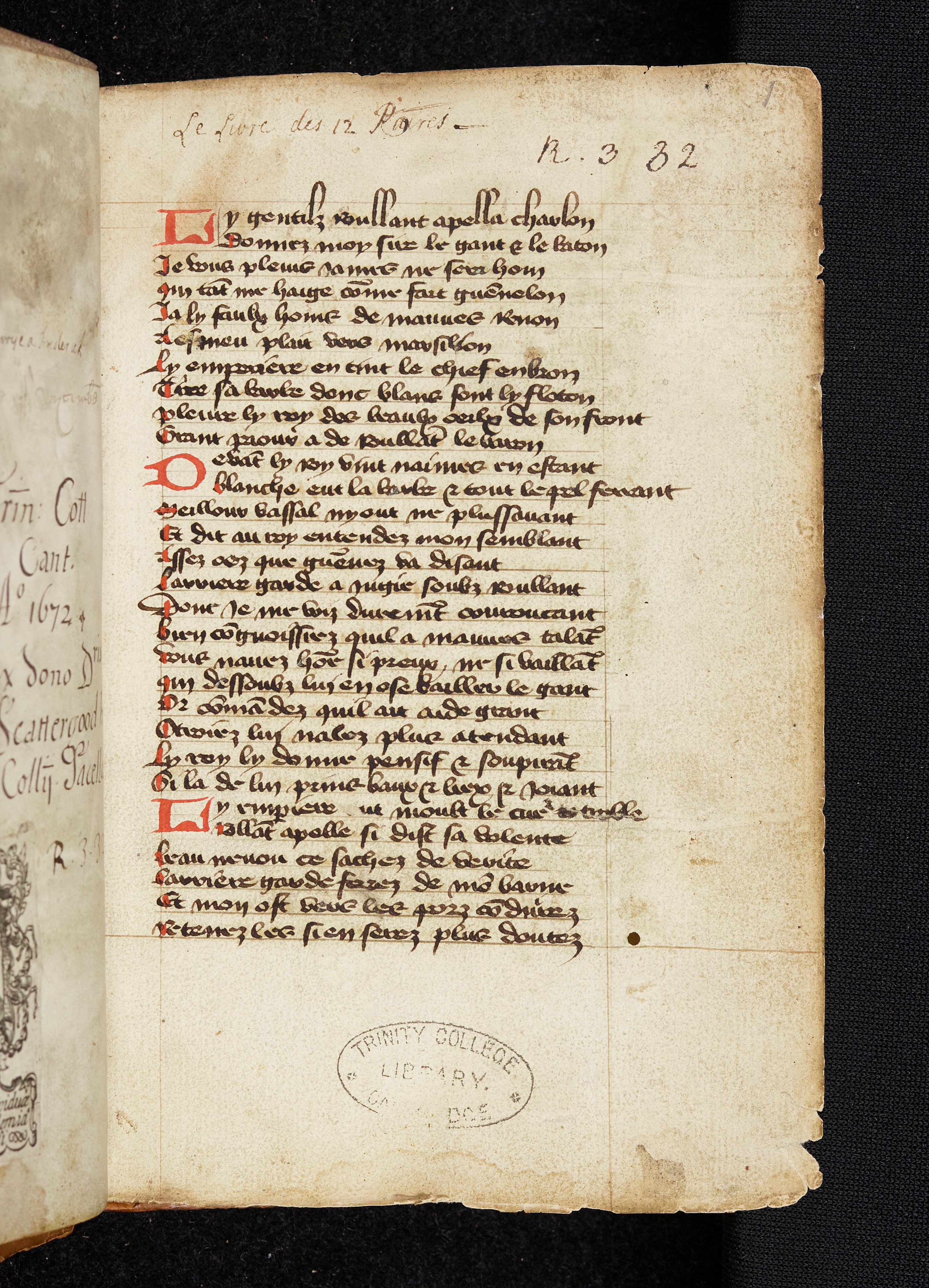}
    \includegraphics[width=0.45\textwidth]{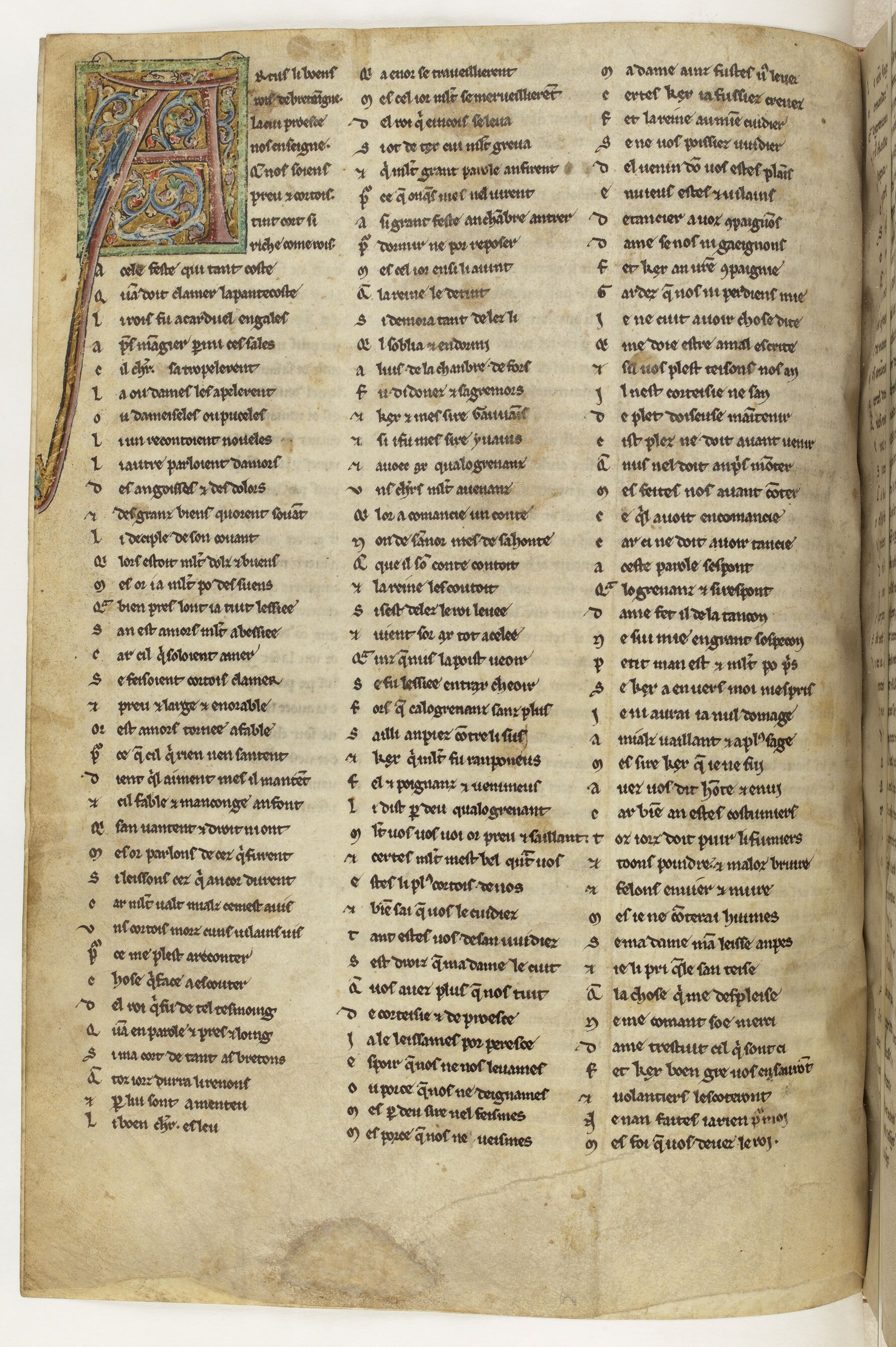}
    \includegraphics[width=0.3\textwidth]{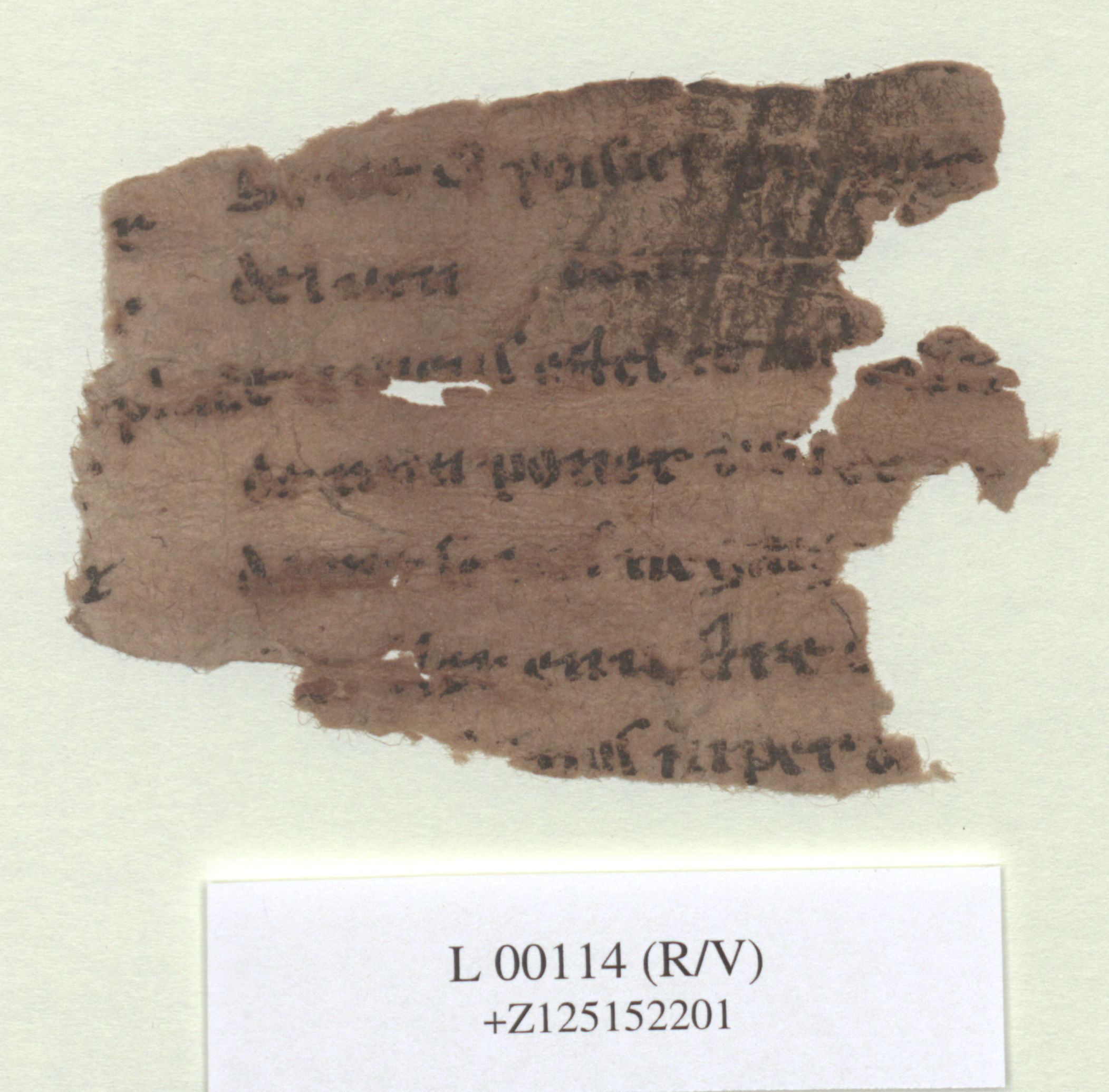}
    \caption{%
    \textbf{Old French chivalric narratives around Europe (years 1100-1500); 1.\,Charlemagne and Roland from the 11th to the 15th century}
    \textbf{A} Beginning of the \textit{Chanson de Roland} (assonanced version, composed around 1090, maybe in Normandy), here in a manuscript copied in England (Oxford?) around 1130 (Oxford, Bodleian Library, Digby 23, Part 2, fol. 1r, - Photo: © Bodleian Libraries, University of Oxford, CC-BY-NC);
    \textbf{B} \textit{Roland} rhymed version, copied in Western France, after 1431
    (Cambdrige, Trinity College Library, R. 3. 32, fol. 1r -- Photo: © University of Cambridge, CC-BY-NC); 
    \textbf{2.\,From Champagne to Egypt, Chrétien de Troyes' \textit{Yvain}},
    \textbf{C} 
    beginning of Chrétien de Troyes' \textit{Yvain ou le Chevalier au lion} (originally composed around 1177 at the court of Marie de Champagne), in a parchment manuscript copied by the scribe Guiot in Provins (Champagne), around 1235 
    (Paris, BnF, fr. 794, fol. 79v; source: gallica.bnf.fr / BnF);
    \textbf{D} fragment on paper of the same text, likely copied in Egypt in the middle of the 13th century 
    (Wien, Österreichische Nationalbibliothek, L 00114 Pap -- Source: Austrian National Library);
    }
    \label{fig:mss}
\end{figure}

We take a well-known case study, that has been subject to renewed attention lately~\cite{kestemont_forgotten_2022}. It concerns the beginnings of Modern European Literature, through the emblematic case of chivalric narratives, the heroic tales of Charlemagne, Arthur or Alexander and their knights, that emerged in French during the 12th century and then spread across Europe. Advantages of this case study include its geographical and chronological reach, as well as its central cultural importance. Indeed, these narratives were widely circulated in an area ranging from England to the Near-East, and subsequently translated in many European languages as diverse as Norse, Irish, Castilian or Middle High German, in a period extending for more than four centuries (\autoref{fig:mss}). Moreover, a large part of them has been well studied, from the 19th century onward, facilitating the collection of information on them such as list of manuscripts, dates and stemmata (\textit{Methods}).    

We ran agent-based simulations of a birth-and-death tradition with a total time frame of $500$~pseudo-years, of which 250 active, and 250 inactive (in which manuscripts can be destroyed, but no longer copied), corresponding roughly, for the first part, to the time from the beginning of the 13th century, seeing the rise of vernacular fiction, to the introduction of the printing press during the Renaissance, and, for the second part, to the time between the Renaissance and the beginning of modern cultural heritage conservation efforts. A pseudo-year is composed of four time-steps in the simulation, each time-step corresponding to $3$ pseudo-months. This was derived using an estimate for the time taken to produce an average $200$~pages manuscript (\textit{Methods}). 
Hence, we simulate 2,000~time-steps, with the second half being ``extinction-only'' ($\forall t \geq 1000, \lambda_t=0$).  

As for the remaining free parameters, namely the base ``birth'' or copy rate, $\lambda$, and ``death'' rate $\mu$, we explore systemically (via discretization) possible pairs of values of these parameters within their range (from $0$ to $1$ in theory, but reduced here to $10^{-3}$ to $10^{-2}$ by field expertise, \textit{Methods}). 

We 
produced heatmaps for a number of relevant observables. 
The approach then consists in identifying, within the parameter space, regions in which the values are consistent either with measured quantities (as is done in the natural sciences) or with estimates for these quantities coming from other, independent and altogether different, models. We have circled in red such regions on the phase diagrams in~Fig.~\ref{fig:phase_diag_1} and~\ref{fig:phase_diag_3}. 

\subsection{Model results on observables}

\begin{figure}[phtb]
 \centering
 \includegraphics[width=\textwidth]{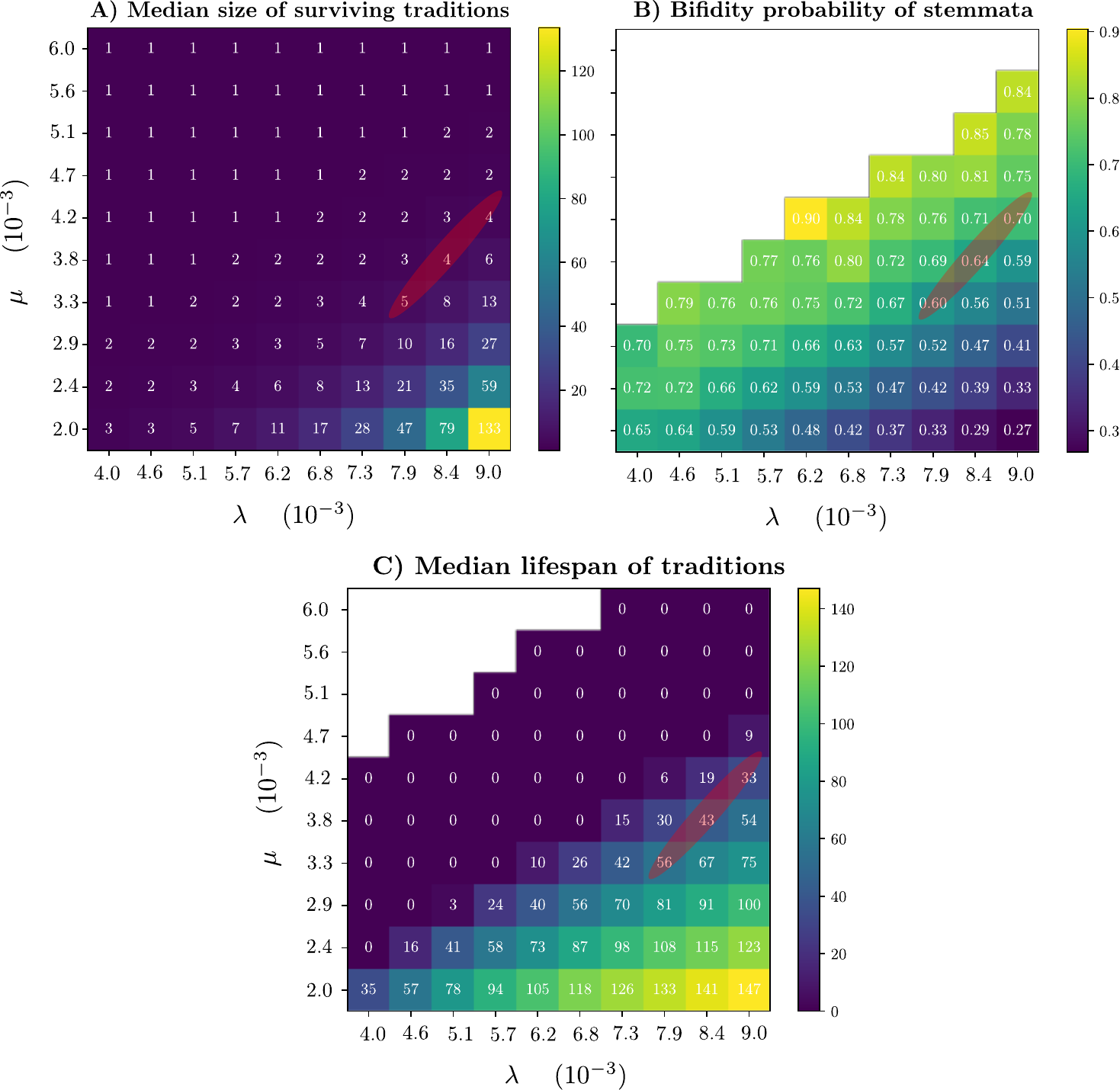}
 
 \caption{\textbf{Heatmaps of model results for some empirically observable data}, namely
  \textbf{A} Median number of surviving manuscripts per tradition (\emph{i.e.} surviving nodes per tree);
  \textbf{B} Proportion of stemmata with root out-degree equal to $2$ (\emph{bifid} stemmata);
  \textbf{C} Average lifespan of a tradition (difference in simulation years between oldest and newest witness birth time). 
  The area circled in red is consistent with historical data
  }
 \label{fig:phase_diag_1}
\end{figure}

These heatmaps show that the results obtained through our simulations are internally consistent in terms of not only population size and lifespan of surviving traditions, but also in terms of bifidity ratio of the resulting trees. In particular, the results obtained for a ratio $\frac{\lambda}{\mu}$ around 2, 
and perhaps more specifically values of 
$\lambda \in [7.5~10^{-3}, 9.5~10^{-3}]$ and $\mu \in [3.2~10^{-3}, 4.4~10^{-3}]$,
are very consistent with the observed properties of medieval chivalric narratives in Old French (\autoref{fig:phase_diag_1}, \textbf{A}, \textbf{B} and \textbf{C}). The proportion of bifid trees in the simulations is slightly lower than the one observed in historical stemmata (60 to 64\% in the simulations; 77\% CI 65\%,86\% in real stemmata).

These results mean that the average manuscript would have been copied $\frac{ \lambda}{\mu}$ times, i.e. two times during their average lifespan. 
The half-life of manuscripts would be $\log(2)/\mu$ time steps, ranging from 60 to 40 years. 
The half-life of all texts, on the other hand, would be close to 500 years (in accordance with survival rates), but only around 30 years for the texts that ultimately did not survive.

\subsection{Loss estimates}

\begin{figure}[phtb]
\centering
\includegraphics[width=\textwidth]{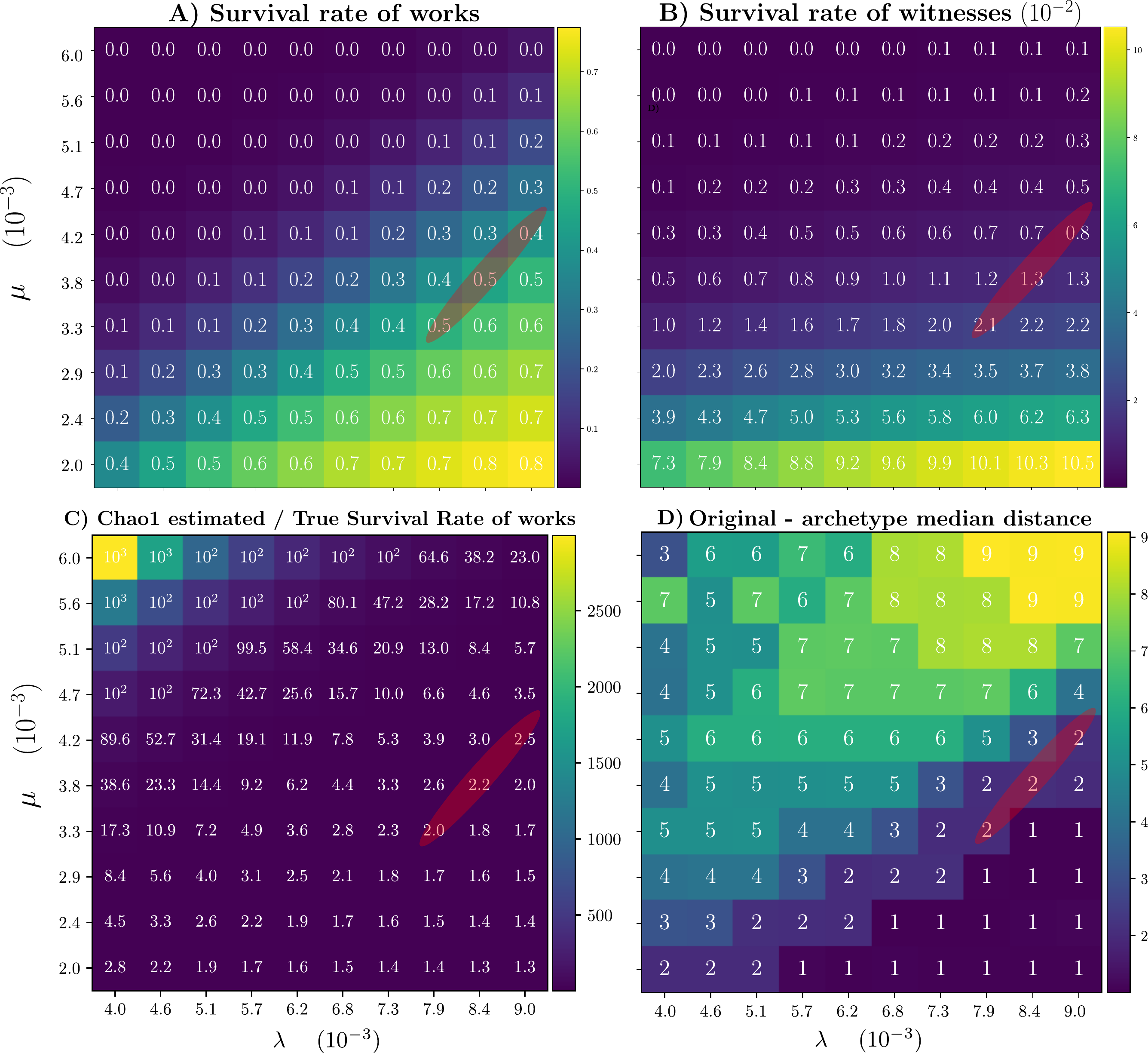}
 \caption{\textbf{Heatmaps of model results for some non empirically observable values}, namely the survival rates computed from birth-death model for \textbf{A} works (\emph{i.e} trees), in the region (in red) with best agreement of the model with the observable, we find values in the range $40\% - 50\%$ and \textbf{B} manuscripts (\emph{i.e.} nodes.) with best agreement around $1\%$; \textbf{C}
 deviations between these survival rates computed exactly on the simulations, and those that would have been estimated using Chao1 on works (estimated to be $2$); 
 \textbf{D} median distance in generations (number of nodes) between the archetype and the original, that is between the root of the full tradition's tree and that of the stemma (best fit : $2$). Here too the area circled in red is the one that proved consistent with historical data on observables (\textit{Methods}).
 }
 \label{fig:phase_diag_3}
\end{figure}

In terms of losses, this yields estimates of 40 to 50\% of survival of works, and around 1\% of survival of witnesses (fig.~\ref{fig:phase_diag_3}, \textbf{A} and \textbf{B}). They are close, yet more pessimistic than those provided by Kestemont et al. \cite{kestemont_forgotten_2022} using the Chao1 species richness estimator \cite{chao1984nonparametric} for the same corpus (resp. 53.5 and 5.4\% as upper bound of survival). It is to be noted that both loss estimates for manuscripts are in the range or close to the global loss rate of non-illustrated manuscripts of 93-97\% estimated by historians based on external evidence (\textit{Methods}).
Interestingly, if the Chao1 estimator had been applied to our simulation results, it would have overestimated survival by a factor of 2  (fig.~\ref{fig:phase_diag_3}, \textbf{C}), showing that there are contexts in which the Chao1 estimator gives an upper bound of survival that can be substantially different from actual survival rates, and lead to underestimate losses\footnote{%
    The fact that our estimates of loss are close to those obtained with a Chao1 estimator on historical data \cite{kestemont_forgotten_2022}, while the application of the same methods to the simulation results would underestimate the loss significantly can be explained by differences in terms of distribution of witnesses per work. Indeed, our model creates data that follows an exponential distribution, while they have, in real data, a power-law type behaviour. 
    Indeed, a large fraction of the total population has a vanishingly small probability of being sampled at any rate in the case of an exponential distribution of witnesses per text, which causes the Chao1 estimator to largely underestimate the size of  the underlying population.
}. 


Beyond the mere probability of survival or extinction, our results also show that the trees that philologists can draw from surviving witnesses would often represent a localised subset of the original tree: the loss of complete branches is apparent in the average distance of 2 generations between the original and the latest common ancestor of the surviving tradition, the archetype (fig.~\ref{fig:phase_diag_3}, \textbf{D}). This means that, in most cases, the original state of the text is not accessible through the preserved tradition, that does not extend beyond a later archetype. It implies also a loss of textual diversity and of the text's evolutionary history, through the disappearance of collateral branches that are entirely missing from the preserved tradition.

\section{Discussion}

\subsection{Solving Bédier's paradox and providing loss estimates}

Adopting a complexity science approach, we have shown how a relatively simple stochastic process that can be computer simulated has the capacity to reproduce the properties observed in historical data for, at least, the textual traditions from medieval French stories about Charlemagne, King Arthur and the fall of Troy. 
Our results are consistent with those obtained through different methodologies, such as those recently published by Kestemont et al.~\cite{kestemont_forgotten_2022}. Our work incorporates previous work by Weizmann and Cisne and provides a much more general framework in which it is possible to refine and tune details from a simple birth-and-death process to a heterogeneous agent-based model calibrated via machine learning methods. This approach allows us to account for population dynamics in time, for loss and production estimates, as well as for certain structural properties of the trees of texts (\textit{stemmata}). In particular we solve Bédier's century-old paradox~\cite{bedier_tradition_1928}, that has been, to this day, at the core of a lasting schism in philological studies.

Indeed, our model shows that a simple, parsimonious stochastic process, without any of the supposed philologists' methodological bias that Bédier assumed, produces a ratio of tree root bifurcation around 60 to 65\% in the range of parameter values consistent with the observed sizes of textual traditions and their average lifespan. This ratio
shows that bifid stemmata are to be expected as the most frequent topology of tree for historical traditions, a fact congruent with existing philological reconstructions  (see Supplementary Table~\ref{models:fig:stemmataCollections}). Yet, it remains slightly lower in the simulations than in historical data for now. Let us emphasise here the fact that further details may be added easily into our model (such as heterogeneity in the copy and destruction rates of manuscripts, meso-scale destruction corresponding to historical events, so-called speciation, etc. --~see~\ref{subsec:call}), that could lead to higher ratios of root bifurcation.

A second important result obtained through our general framework is a set of refined estimates for the loss of texts and manuscripts across centuries --~these are the first estimates based on computer simulations of artificial manuscript populations. Consistent with the upper range of loss estimates provided by external historical evidence (see Supplementary Table~\ref{tab:death_rates_chivalric}), our results bring new insights into the vast continent of lost cultures and establish a link between the immaterial loss of cultural diversity (both qualitatively and quantitatively) and the disappearance of entire branches from the trees of texts, as well as a relatively high ratio (up to 60\%) of fully missing trees. This makes evident how partial and fragmented our perception of past cultures may be.

Further, our results show the importance of taking into account diachrony and the evolutionary nature of text transmission. In addition to a better understanding of the process of text transmission itself, these results tend to indicate how a purely synchronous approach, as the one with ecodiversity methods, could lead to estimates relatively far from actual losses in time, and would not allow us to fully grasp the way in which loss is distributed. This is an important aspect, as the loss of entire branches (let alone entire trees) has a very different impact in terms of observed diversity than the mere loss of individuals distributed in surviving branches.   

\subsection{A call for evolutionary modelling of text transmission\label{subsec:call}}

\begin{figure}[h]
    \centering
    \includegraphics[width = \textwidth]{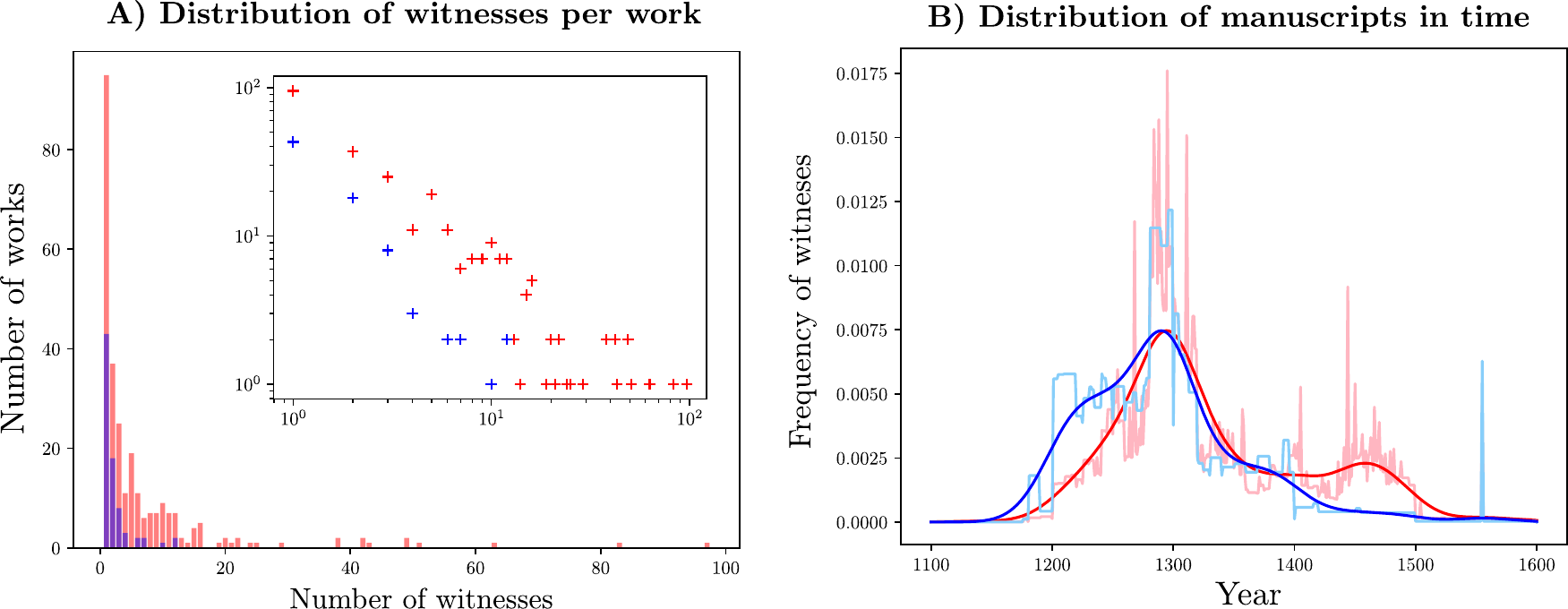}
    \caption{%
    \textbf{A} The distribution of the number of witnesses per work shows a Pareto-like, heavy-tail behaviour which does not correspond to the exponential distribution produced by our first level model (\textit{Methods}). \textbf{B} Distribution of manuscripts production date for historical data. In these figures, surviving witnesses are shown in red, and surviving fragments of lost manuscripts in blue.
    }
    \label{fig:distribs}
\end{figure}

\begin{table}[htbp]
    \centering
    \caption{Comparison of various topological properties for real stemmata and simulated ones in the region of best agreement in the parameter space. Upper and lower bounds for our estimations on real world data are computed from the 10\% confidence interval of the bootstrap distribution over our corpus of 117 stemmata, with 100 resamplings. The range indicated under \emph{Simulations} corresponds to the min/max values in the region of best agreement in parameter space. The \emph{direct witnesses connections} percentages are the proportion of all edges in stemmata that are filiations between surviving manuscripts. The degree distributions $\mathbb{P}(\text{degree} = k)$ are restricted to internal (non-leaf) nodes. As for the definition of the imbalance index $i_3$, see \textit{Methods}.}
    \begin{tabular}{||l|c|c||}
        \hline
        Property & Real stemmata  & Simulations \\
        \hline\hline
        Bifid tree proportion & \textbf{77 \%} $\begin{array}{c} 86\% \\ 65\%\end{array}$ & \textbf{60 \% -- 64 \%}\\
        \hline 
        Direct witnesses connections & \textbf{2 \%} $\begin{array}{c} 4 \% \\ 1 \% \end{array}$ &  \textbf{1 \% -- 4.5 \%}\\
        \hline
        $\mathbb{P}(\text{degree} = 2)$ & \textbf{0.81} $\begin{array}{c} 0.87 \\ 0.76 \end{array}$ & \textbf{0.71 -- 0.78}\\
        \hline
        $\mathbb{P}(\text{degree} = 3)$ & \textbf{0.11} $\begin{array}{c} 0.16 \\ 0.08 \end{array}$ & \textbf{0.14 - 0.15}\\
        \hline
        $i_3$ & \textbf{0.93} $\begin{array}{c} 0.96 \\ 0.89 \end{array}$ & \textbf{0.80 -- 0.85}\\
        \hline
    \end{tabular}
    \label{tab:main:tree_properties}
\end{table}

Our results clearly show how much of the transmission, survival and extinction of texts can be explained by drift, and how some of the salient properties of textual traditions, in terms of chronology, sizes of traditions or topological features of stemmata, may in large part be explained by the role of chance. Yet, they also point to the existence of other potential factors that need to be accounted for, such as the interrelation between texts and the possible role of selection and extrinsic factors.

Indeed, considering texts as independent entities does not fully reflect the ways texts were created, very often by a process of derivation from preexisting texts, by translating, adapting, or rewriting existing material. This might be apparent in the way in which our model gives an exponential distribution for the number of witnesses per work, while real data appear to follow a Pareto-like heavy-tail behaviour~(\autoref{fig:distribs}, \textbf{A}). The type of distribution observed is an intrinsic property of the qualitative dynamics of the model, and this discrepancy suggest that some important aspects of the dynamics of actual text diffusion are not captured --~this does not come as a surprise, since the complex systems methodology voluntarily starts from the most basic model possible, before climbing one step at a time up the ladder toward more sophisticated models, in order to see which ingredients suffice to produce each observed property. In our case, accounting for the apparition of new texts from existing ones, through a process of speciation, might be enough to capture the heavy-tail distribution of the number of witnesses in historical data. Indeed, a famous and widely used model including speciation, the Yule process~\cite{yule_1925}, is known to produce power-law type distributions, and was originally aimed at understanding the power-law distribution of species per genus.

Additionally, our results point to the importance of extrinsic factors and perhaps also of selection. Indeed, in our simulations, the distribution in time of witnesses grows exponentially, but historical data exhibit other kinds of variation (\autoref{fig:distribs}, \textbf{B}), with ups and downs reflecting literary trends or external events, such as the Black Plague of 1346-1353, the Hundred Years' War or more generally the economic and demographic crisis of the late Middle Ages \cite{camps_make_2023}. Models including selectivity, or accounting for extrinsic factors and non-random drivers of extinction such as climate change in biology~\cite{yessoufou_reconsidering_2016}, 
could help account for these discrepancies. 
Moreover, if our simulations solve Bédier's paradox on tree root-bifurcation,
they still produce trees that are relatively less imbalanced than actual stemmata (\autoref{tab:main:tree_properties}, see also section~\ref{sec:imbalance}). This is a problem that is also observed in evolutionary biology~\cite{mooers_inferring_1997,aldous_2001, khurana_2023}, and constitutes a further indication that drift alone cannot fully explain the process of text transmission. Let us emphasise once more that our general framework is designed to accommodate the more detailed models mentioned in this discussion, and that this is precisely one of its main advantages.

Another point that requires research attention is to account for evidence not strictly falling into the categories of survival or loss: in particular, a large number of fragments (a few leaves of otherwise lost manuscripts) have been preserved across time. They may not be considered cases of survival, yet they give empirical proof of the existence of lost manuscripts, in a way similar to fossils in paleodemography\footnote{%
   There are also more pragmatic reasons for not including fragments in this research: very often, it is hard to assess if different fragments of a few leaves originally came or not from a unique now lost manuscript; moreover, due to their size, they usually offer very limited usable information in terms of date, origin, or even identity of the text they contain. Worse, due to their fragile nature, they still are often not identified or lost by curatorial institutions, limiting the quantity of information available to us today.
}.
Yet, existing evidence still points to the fact that fragments behave similarly as surviving witnesses in terms of distribution, and do not radically challenge results obtained only on surviving witnesses, while being slightly older in terms of date, reflecting the cumulative damage in time affecting manuscripts (\autoref{fig:distribs}, \textbf{A} and \textbf{B}).

Finally, the generality of our framework and the models it can accommodate makes it applicable not only to medieval texts, but to any type of cultural transmission, at least in written form. 
Further investigations should include the broadest possible range of cases, starting with Western Medieval and Antique texts, but preferably also encompassing cultural productions from very different time periods and geographical areas. In this sense, our work calls for the emergence of a new strand of research dedicated to evolutionary modelling of cultural transmission.

\end{bibunit}

\clearpage

\begin{bibunit}
\section{Methods} \label{sec:methods}

\subsection{A dynamic model of text transmission}

The continuous-time Markov process that we consider in this work is a multi-agent generalization of the birth-and-death process. The state of the system at any time is described by the number $M_t$ of agents (representing manuscripts) generated up to time $t$ as well as the states $\{s_i\}_{0 \leq i \leq M_t}$ of these agents with $s_i \in \{\mathbf{living}, \mathbf{dead}\}$. For each living agent $i$ at time $t$, two transitions can occur during instant $[t,t+\mathrm{d}t]$:
\begin{itemize}
    \item 
    A birth event with rate $\lambda$, such that with probability $\lambda~ \mathrm{d}t$ a new agent (originating from $i$) is added to the system with state $s_{M_t + 1} = \mathbf{living}$,
    \item
    A death event with rate $\mu$ such that the state of $i$ switches to $s_i = \mathbf{dead}$ with probability $\mu~\mathrm{d} t$.
\end{itemize}
Starting with $M_{t=0} = 1$ with a single living agent, we split the time evolution of the process into two phases:
\begin{itemize}
    \item 
    An active phase for $0 \leq t \leq t_a$ where both $\lambda$ and $\mu$ are finite and constant
    \item
    A decimation or pure-death phase for $t_a \leq t \leq t_a + t_e$ where $\lambda =0$.
\end{itemize}
Moreover, we keep track of the parent-offspring relationship, so that the state of the system can be alternatively described by a directed tree with nodes standing for agents, where an edge $(i,j)$ is drawn when agent $j$ was generated by a birth event triggered by $i$ (\emph{i.e.} $j$ was copied from $i$). We call a single realisation of this process a \emph{manuscript tradition}.

\subsubsection{Pure birth-death processes with fixed or variable rates} \label{sec:exact_computations}

The probability distribution of the number of surviving manuscripts in a tradition, as well as the survival rates of manuscripts and traditions can be computed analytically for the model with active and decimation phase. With $t_a$ the duration of the active phase and $\overline{n}_t = e^{(\lambda - \mu) t}$ the average number of living manuscripts (for $t \leq t_a$), the probability that a given tradition has $n$ surviving witnesses at $t_a$ is given by~\cite{kendall1948generalized}
\begin{equation*}
    P_n (t_a) = \left\{ \begin{array}{ll}
    \frac{\mu}{\lambda} \eta_{t_a} & \;\; \text{if} \; n = 0\\
    (1-\frac{\mu}{\lambda} \eta_{t_a}) (1 - \eta_{t_a}) \eta_{t_a}^{n-1} & \;\; \text{if} \; n \geq 1
    \end{array} \right.\;,
\end{equation*}
where
\begin{equation*}
\eta_t = \frac{\lambda (\overline{n}_t -1)}{\lambda \overline{n}_t - \mu}\;.
\end{equation*}
Denoting now by $t_e$ the duration of the decimation phase, the probability that the final population of living manuscript is $n$ given that there were $k$ extant manuscripts at time $t_a$ is 
\begin{equation*}
    P(N(t_e) = n \mid N(t_a) = k ) = \binom{k}{k-n} (1-e^{ - \mu t_e})^{k-n}
\end{equation*}
so that the final probability for the number of surviving witnesses for $n \geq 1$ is
\begin{align*}
    P_{n} (t_e + t_a) & = \sum_{k = n}^\infty \binom{k}{k-n} \left(1-e^{ - \mu t_e} \right)^{k-n} P_k (t_a)\\
    & = \left(1 - \frac{\mu}{\lambda} \eta_{t_a} \right) \frac{1 - e^{-\mu t_e}}{\eta_{t_a}} \left(\frac{1}{\eta_{t_a}} (1-\eta_{t_a})(e^{\mu t_e} -1) \right)^{-n} \;.
\end{align*}
while for $n=0$ the extinction rate of traditions -- or equivalently the survival rate of works $s_\text{Works} = 1 - P_0(t_a + t_e)$, is
\begin{equation*}
    P_0(t_a + t_e) = \frac{\mu}{\lambda} \eta_{t_a} + \left(1- \frac{\mu}{\lambda} \eta_{t_a} \right) \left(\frac{1}{\eta_{t_a}} - 1\right) \frac{\eta_{t_a} (1 - e^{\mu t_e})}{1 - \eta_{t_a} (1 - e^{\mu t_e})} \;.
\end{equation*}
The number of surviving witnesses follows a geometric distribution for $n \geq 1$, from which one can compute the median number of witnesses of surviving traditions plotted in Fig.~\ref{fig:phase_diag_1}.

It can also be shown \cite{kendall1948generalized} that the average cumulative population of a given tradition, \emph{i.e.} the total number of manuscripts produced during the active phase, writes
\begin{equation*}
    \overline{M}_{t_a} = 1 + \frac{\lambda}{\lambda - \mu} \left(e^{(\lambda - \mu) t_a} -1 \right) \;,
\end{equation*}
while the average number of living manuscripts at $t_a + t_e$ is
\begin{equation*}
    \overline{N}_{t_a + t_e} = e^{(\lambda - \mu) t_a - \mu t_e}
\end{equation*}
so that the survival rate of manuscript over all traditions writes
\begin{equation*}
    s_\text{Man} = \frac{\overline{N}_{t_a + t_e}}{\overline{M}_{t_a}}\;.
\end{equation*}
These results are plotted on figures~\ref{fig:phase_diag_1} and~\ref{fig:phase_diag_3}.

\subsubsection{From simulated trees to stemmata}

Once the full manuscript tradition originating from a given original is generated, one has to construct the corresponding stemma in order to allow for comparison with real philological data. This stemma should contain only surviving nodes of the full tree along with the minimal number of dead nodes needed to keep track of the genealogical relationship between those. In particular all terminal nodes (leaves) of the tree have to be surviving witnesses, and all internal (non-leaves) nodes need to have at least two direct children. Indeed, chains of  non-branching (parents of single child) hypothetical ancestors cannot be inferred from the comparative method, and thus shouldn't appear on a stemma by application of the parsimony principle.

\begin{figure}
    \centering
    \includegraphics[width=\textwidth]{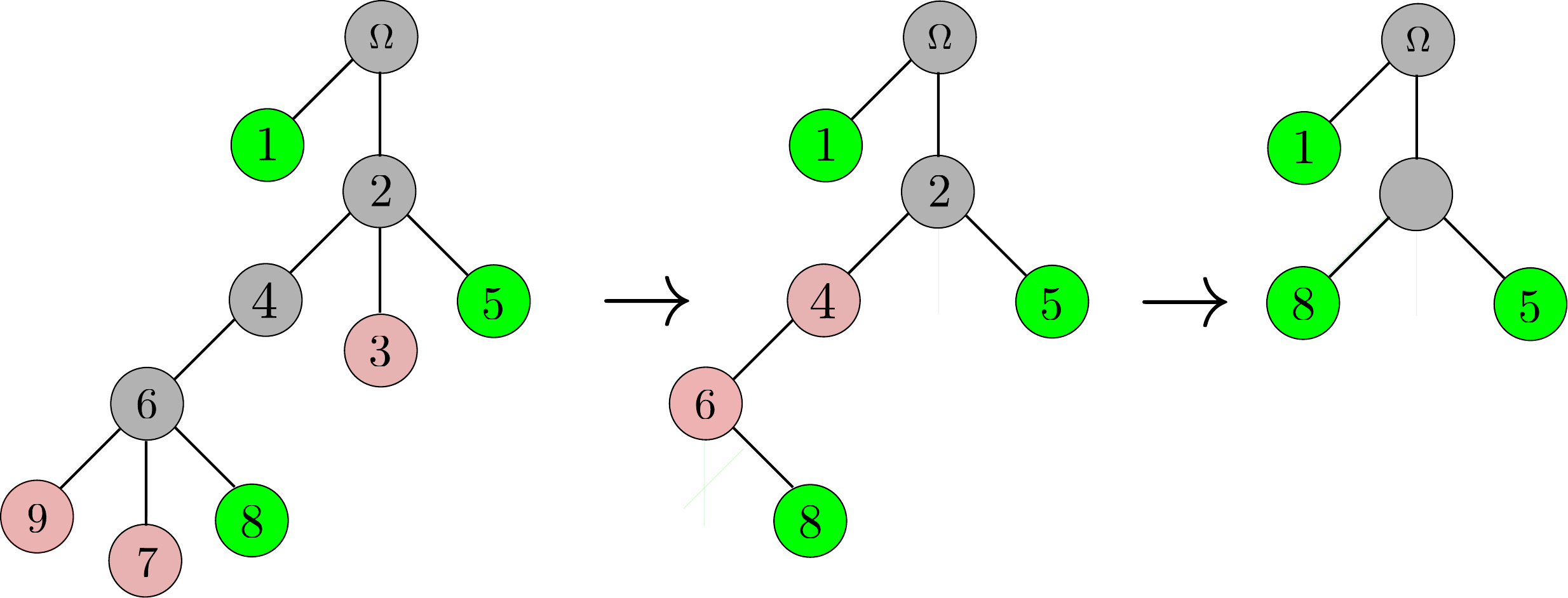}
    \caption{Steps of the algorithmic construction of the stemma from the full tree of a simulated tradition with original $\Omega$, surviving manuscripts in green and lost ones in grey and red: removal of dead branches (left), removal of non-branching lost manuscripts (middle), final stemma (right).}
    \label{fig:stemma_reconstruction}
\end{figure}

The stemma is then constructed from the full tree by first recursively removing dead leaves until all branches have only surviving manuscripts as terminal nodes, then removing chains of non-branching dead nodes, until all remaining dead nodes have at least out-degree two (see figure~\ref{fig:stemma_reconstruction}).

The resulting reduced tree then corresponds to the ideal (correct) stemma that would be reconstructed \emph{a posteriori} by a philologist based on the systematic comparison of the witnesses of the tradition.

\subsection{Simulations parameter values}

\subsubsection{Time-step and total duration}

The base time-step of the model is the time necessary to copy a manuscript.  This is of course dependent on the length of the text and the speed of copyists. Other factors may have played a role, such as the care and quality of execution, the script type, the presence or not of ornamentation implying the interventions of painters after the end of the copy itself, or the existence of modes of serial productions such as the \textit{pecia} system, which allowed several scribes to work simultaneously on copying a single exemplar.

The speed of scribes is known to vary a lot, from less than $1$ to $10$~leaves a day,
while the total time taken to copy a volume often counted in months, often between 3 or 6 months, and in some rare cases in years (up to 12 years)~\cite{overgaauw_fast_1995}. 

A survey of observable copy rates of professional scribes (see Supplementary Table~\ref{tab:copy_time}), that are expected to have been most of the time the producers of vernacular literary manuscripts in the 13th to 15th centuries, gives an average rythm of between 2 and 3 folio per day (f/d), with a mean rate of  of 2.76 f/d
, equivalent to a duration of 72 days for a volume of 200 leaves, between two or three  months.

We wish to model the transmission of medieval chivalric texts, that originated mostly in the years 1150-1350 and went out of fashion during the Renaissance, and then knew a period of inactive life (destruction, but almost no copy) until the beginning of preservation efforts during the Industrial Revolution. We therefore aim for a total duration of the simulation of 500 pseudo-years, split in two between an active period of 250 years and an equal inactive period of 250 years, each then being composed of 1000 simulation time-steps of three pseudo-months.

\subsubsection{Copy (or birth) and loss (or death) rates}

Historical and philological knowledge of loss rates is very scarce and elusive. However, it is possible to approach them through different types of external evidence, such as data from ancient library catalogues~\cite{buringh_medieval_2010}, inventories or wills, as well as using intertextual evidence, such as allusions or references to unknown texts~\cite{wilson_r_m_lost_1952,bardon_litterature_1952}  (see Supplementary Tables~\ref{tab:death_rates_chivalric} and \ref{tab:death_rates_other} for more details). 
    
    \paragraph{Ancient catalogues.} Using data based on a large collection of historical catalogues of manuscripts, 
    Buringh \cite{buringh_medieval_2010} provides estimates for the Latin West, with a geometric mean of loss around -25\% per century, with variations from -11\% in the 9th to -32\% in the 14th and 15th centuries (with local variations between medieval institutions from –3\% to –71\% per century). The global loss rate for non-illustrated manuscripts of several well known collections has been estimated around 93-97\% \cite{kestemont_estimating_2020,neddermeyer_von_1998,wijsman_luxury_2010,oostrom_stemmen_2013}. But estimations based on well known institutional collections, from which some manuscripts are known to have survived, are potentially biased. Trying to account for fully lost libraries, Buringh \cite{buringh_medieval_2010} is compelled to revise his estimates higher, to -25\% by century until the 12th, up to -43\% in the 15th. 
    
    Moreover, generally speaking, using catalogues leads to several biases that can lead to an underestimation of the loss rates: 
    \begin{enumerate}
        \item medieval catalogues are produced mostly by institutions, and among them mostly ecclesiastical institutions; 
        \item they concern mostly Latin religious texts, that are presumed to have been less decimated than vernacular leisure literature; 
        \item catalogues reveal the existence of a preservation effort; 
        \item catalogues do not necessarily record all books, only the ones that are deemed most worthy of description and preservation. 
    \end{enumerate}
    
    All this can lead to underestimate the loss rate, and is also tributary to the fate of ecclesiastical institutions. In France, where many of them were maintained until the Revolution, and their assets then seized and included in national collections, the loss rate estimated from  Buringh's data collection \cite{buringh_medieval_2010} is 72\%. In England, where the Dissolution caused the scattering and destruction of many ecclesiastical collections, the loss rate is 91\%. In addition, the few data that we have concerning non ecclesiastical collections, that included more vernacular literature, seem to confirm this bias: the loss rate for the Royal Library of the Louvres, in France, is of 92\% \cite{buringh_medieval_2010}, while, for the few examples of aristocratic libraries recorded by Buringh, it reaches 100\% loss.
    
    In addition to the institutional nature of the library preserving them or the nature of the text, the monetary value of books causes also important variations in survival rates~\cite{bozzolo_pour_1980}, in particular when they are richly decorated with miniature and other paintings. For instance, concerning the luxury collection of the dukes of Burgundy, Wijsman \cite{wijsman_luxury_2010} notes both a low destruction rate (41\% for the books in the 1487 catalogue), but also notes that those catalogues tended to exclude books of lower values.

    \paragraph{Print runs of early prints.}
    For incunabula, using editions whose original number of copies made is known, it is possible to gather loss estimates by counting known surviving exemplars in public or private collections: doing so for Venetian incunabula, Trovato \cite{trovato_everything_2014} finds very variable loss rates according to textual and material typology, from 73\% for the \textit{Decretales} printed on parchment to 99.3\% for more popular chivalrous literature (\textit{Orlando furioso} for instance). This shows the importance both of variation in time and space, and of textual contents and material typology. A more general estimate for 15h century prints yields a loss rate of 95.7\% for Europe~\cite{neddermeyer_moglichkeiten_1996}.
    
    \paragraph{Other estimates.}
    In some extreme cases, loss can be very close to 100\%, for reasons that may combine the fragility of the document form, lack of consideration for the documents or large scale historical events such as political instability, invasions or major cultural changes; examples are provided by cases as different as the Merovingian royal diplomas on papyrus or the Lombard royal charters \cite{ganz_charters_1990}, the Mayan (pre-colombian) manuscripts 
    or medieval notarial acts \cite{holtz_uberlieferungs-_2001}.
    Production estimates have also been attempted on the basis of the quantity of
    sealing wax acquired by a given producer (a chancellery for instance \cite{bautier_introduction_1978}).
    More founded loss estimates have also been gathered by counting how many of the acts mentioned in imperial or royal registers are kept in original or consigned in the archives of the recipients: this gives a loss rate of originals varying from 80\% (acts from the emperor Charles IV in 1360-1361) to 90\% for the acts from Louis X of France, increasing to 99\% for the judgements rendered by his  Parliament, suggesting here as well a massive effect of typological variation \cite{holtz_uberlieferungs-_2001,canteaut_quantifier_2020}, resulting in very strong biases in the body of  documents available to us.

    \paragraph{Order of magnitude for copy and loss rates in the simulations.}
    Most general estimates give a total loss that is superior to 90\%.  For chivalric texts, if we look at Trovato's estimates (potentially more reliable, because based on editions whose original number of copies is known), we have a total loss rate superior to 99\% (99.3\% on average). 
    For our simulation needs, on this basis, we get a survival rate whose order of magnitude is between 0,1\% and 10\% in 500 years (2000 steps in our model).
    From this we can deduce a step loss rate for a given total survival rate. For instance, for 1\% survival rate,  
    $(1-\mu)^{2000} = 0.01$, which simplifies to $\mu = 0.0023$. So we retain values of $\mu$ between $10^{-3}$ to $10^{-2}$. 
    Given that books could not have been produced order of magnitudes faster or slower than they were destroyed (or we would either drown in medieval manuscripts or have none), we explore the same range for $\lambda$. 
    
\subsection{Tree imbalance} \label{sec:imbalance}

Traditional philological studies have focused on the proportion of bifidity in stemmata collection. Yet this measurement is relatively restricted in scope and does not reflect more general topological properties of the trees, and in particular their tendency to display imbalance at all levels. The characterization of imbalance is the object of a considerable literature in computational phylogenetics \cite{aldous_2001,mooers_inferring_1997}, and a number of metrics have been proposed to quantify this property \cite{fischer_et_al_2023}. In this work we chose to focus on a single imbalance index for the stemmata, based on the subtrees generated by subsets of leaves representing witnesses of a tradition.

\begin{figure}
    \centering
    \includegraphics[width=0.5\textwidth]{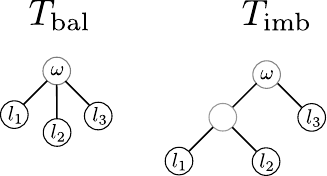}
    \caption{The two possible topologies of a 3-leaves tree with only branching internal nodes}
    \label{fig:3-subtrees}
\end{figure}

There are only two possible topologies for a rooted tree with three leaves (forbidding internal nodes with a single child), one maximally imbalanced denoted by $T_\mathrm{imb}$, corresponding to the simplest "caterpillar tree" in the terminology of \cite{fischer_et_al_2023}, and one perfectly balanced $T_\mathrm{bal}$ with root node having three children (see figure~\ref{fig:3-subtrees}). Given a stemma with $N_l$ leaves, we construct the subtrees generated by all subsets of tree leaves $\{l_1,l_2,l_3\}$ which then fall into one of the two previously defined cases. Denoting by $n_\mathrm{imb}$ the number of generated subtrees having a $T_\mathrm{imb}$ structure, the imbalance index $i_3$ of the stemma is defined as the proportion of imbalanced subtrees
\[i_3 = \frac{n_\text{imb}}{\binom{N_l}{3}}\]
This specific index, rather than other summary indices used to measure imbalance of tree topologies seems particularly fit to the study of stemmata as the construction of subtrees generated by witnesses is in itself an elementary emulation of the comparative method in philology, thus making the index readily interpretable. Besides, it allows for a rather scale-independent estimation of imbalance, as the root $\omega$ of a given subtree, or \emph{hyparchetype} of the corresponding set of witnesses, ranges from low lying internal nodes to the actual archetype of the whole stemma when different subsets of witnesses are considered.

The value of this index on real stemmata (see table~\ref{tab:main:tree_properties}) is found to be equal to 
$0.93~(\pm 0.03)$, a value significantly higher than what is found in simulated stemmata for a wide range of parameters. Remarkably, the fact that empirical evolutionary data display a higher level of imbalance than cladograms generated by constant-rate birth-and-death models is also observed in biological phylogenetics \cite{khurana_2023}.

\subsection{Observables of historical traditions}

Data about the works and witnesses of medieval texts were collected from secondary literature and existing databases \cite{camps_`chanson_2016,martina_produzione_2018,vitale-brovarone_diffusion_2006,irht_jonas1993,brun_arlima_2005,kestemont_forgotten_2022} and reviewed using editions of the texts when they were available. Information on the stemmata is based on a restriction to the relevant genres of the data provided by the open-source collection OpenStemmata \cite{camps_open_2021,camps_open_2021b}. All used data (and relevant code) is provided as supplementary information to this article.

\subsubsection{Corpus and data collection}

The corpus is composed of longer narrative works in Old French (i.e. composed before c.~1340), namely \textit{chansons de geste} (epics) and \textit{romans} (romances) in verse and prose. They are traditionally divided in three main thematic areas,  or ``matters'': matter of France (stories about Charlemagne, his lineage and his peers); matter of Britain (stories about Tristan or King Arthur and his knights); matter of ``Rome'' (stories inspired by Antiquity, and dealing with Alexander the Great, with the Fall of Troy and its consequences, or with Theban and Oriental myths). In addition from texts dealing with these matters, texts of the matter of England (stories of noble Anglo-Norman or English families) and of the Crusades were also included, as well as standalone adventures or courtly romances, containing similar narratives about knights and lovers. Shorter narrative forms, such as \textit{lais}, \textit{fabliaux} were excluded, as well as romances not fitting in these thematic categories (\textit{roman de Renart}, and allegoric romances such as the \textit{Roman de la Rose}). 

The list of texts was established by crossing lists existing in the literature \cite{camps_`chanson_2016,martina_produzione_2018,vitale-brovarone_diffusion_2006,hasenohr_dictionnaire_1994}, with existing databases \cite{kestemont_forgotten_2022} and online reference repositories \cite{irht_jonas1993,brun_arlima_2005}. When available, existing editions of individual texts were consulted, to verify the information and to collect existing stemmata. When no stemma was found in editions, scientific articles listed in the repositories were browsed to find them.  

The notion of texts is relatively fluid in manuscript transmission, where every copy introduces innovations, that can range from simple copying errors to complete rewriting. Very often, decision had to be made whether two versions of a given story constituted two versions of the same text or two distinct texts. To establish this, in ambiguous cases were information in repositories and reference works did not fully coincide, the following definition of text and set of criteria were used:
a text is taken to be the formulation of a given story in human language, whose literary form and style can be detected and whose creation can be attributed to one or more individuals. Following this, two versions of the same story were taken to be different texts when:
\begin{itemize}
    \item they differ in terms of language (e.g., Continental French vs. Anglo-French);
    \item they differ in terms of form: prose vs. verse; octosyllabic versus decasyllabic verses; assonance vs. full rhyme;
    \item they differ in authorship (a different individual is associated with their creation);
    \item they substantially differ in terms of story, or their texts cannot be aligned (i.e. there are no significantly matching text portions).
\end{itemize}

For each text, the following information was collected, in addition to author and title:
\begin{itemize}
    \item date of creation (usually estimated based on historical, linguistic and literary evidence);
    \item witnesses;
    \item stemma.
\end{itemize}

The stemmata were encoded in DOT format, and submitted to an open repository using their set of conventions for representing surviving and lost nodes, as well as direct or lateral transmission and uncertainty \cite{camps_open_2021b}.
In addition, the following information was collected about the witnesses:

\begin{itemize}
    \item date of creation (usually estimated based on material, linguistic and paleographic evidence, sometimes known through declarations by the scribe, known as ``colophons'');
    \item curatorial information: institution of preservation and shelfmark;
    \item status (complete, mutilated or fragmentary).
\end{itemize}

\subsubsection{Dates and date ranges}

Most of the times, the dates of the texts or of the witnesses are not known precisely, but estimated. Estimates made by experts are usually expressed in relatively vague form, in prose. We opted to convert them to explicit ranges, following this set of rules:
\begin{description}
\item[Exact dates] exact dates, sometimes given by the scribe of the manuscript in the colophon, are followed; 
\item[Explicit Range] the range given is kept;
\item[Non explicit range] a range of 20 years is used, bounded or centred according to the information given;
\item[Cumulative ranges] both ranges are included. When the ranges given are not continuous, the in-between period is added;
\item[Fuzzy date/range] a period of 10 years (5 before, 5 after) is added to the date or range expressed ; 
\item[Single delimiter pseudo-range] i.e., terminus ad quem/post quem without explicit range) are treated similarly as fuzzy dates;
\item[Exotic modifiers] such as ``at the latest'' are ignored for now; 
\item[Competing dating] when two dating are given, the most precise is used.
\end{description}

\begin{table}[h]
    \centering
        \caption{\textbf{Example of conversion of prose dating to ranges}}
    \label{tab:methods:dates}
    \begin{tabular}{lr}
    \textbf{Prose} & \textbf{Range}\\ \hline\hline
    \multicolumn{2}{c}{\textit{Exact dates}}\\ \hline
      17th may 1423 & 1423 \\ \hline
      \multicolumn{2}{c}{\textit{Explicit range}}\\ \hline
       12th century & 1101-1200 \\ 
       2nd half of the 12th century & 1151-1200 \\ 
       14th century, after 1339 & 1340-1400\\\hline
      \multicolumn{2}{c}{\textit{Non explicit range}}\\ \hline
       end of the 12th century & 1181-1200 \\ 
       middle of the 12th century & 1141-1160\\ 
       first decades of the 14th century & 1301-1320\\\hline
      \multicolumn{2}{c}{\textit{Cumulative ranges}}\\ \hline
       end of the 12th or beg. of the 13th c. & 1181-1220\\
       beg. or 3rd quarter of the 15th c. & 1401-1475 \\\hline
      \multicolumn{2}{c}{\textit{Fuzzy date or range}}\\ \hline
       circa 1240 & 1236-1245\\
       around the end of the 12th c. & 1176-1205\\\hline
      \multicolumn{2}{c}{\textit{Single delimiter pseudo-range}}\\ \hline
       after 1460 & 1461-1470\\
       before 1460 & 1451-1460\\ \hline
      \multicolumn{2}{c}{\textit{Exotic modifiers (\textit{ignored})}}\\ \hline
      beg. of the 13th c. at the latest & 1201-1220\\\hline
      \multicolumn{2}{c}{\textit{Competing dating}}\\ \hline
      2nd quarter of the 13th c., c.~1240 & 1236-1245
    \end{tabular}
\end{table}

Example of conversions are given in \autoref{tab:methods:dates}.

\subsubsection{Extraction of summary statistics from data and parameter estimation}
\label{sec:model_based_estimation}

In order to fit the model to real world data, a subset of observables on manuscript traditions were selected for comparison with the model's results:

\begin{itemize}
    \item 
    The proportions $f_k$ of surviving traditions consisting of $k$ witnesses for $1 \leq k \leq 3$; 
    \item
    The median lifespan of a tradition defined as the difference in age between the oldest and newest witness of a tradition. Since our model does not account for historical extrinsic factors on the variation of manuscripts and works production rates, we expect this observable to be invariant enough by translation in time to allow for meaningful comparison with the model;
    \item
    The proportion of trees with root-degree 2 (see section~\ref{sec:philo_for_imbal} on the relevance of this specific feature);
    \item
    The proportion of internal nodes of degree 2 as an additional topological feature on stemmata.
\end{itemize}
These summary statistics can be readily extracted from simulations, however some additional interpretative work is needed on empirical data.

Regarding temporal data, in order to convert collected evidence into distributions of creation date of manuscripts and lifespan of traditions, we assumed a uniform probability distribution for the birth time of a manuscript whenever its dating was given as an interval. Thus for a manuscript with creation date estimated to be within an interval $[y_\text{l}, y_\text{u}]$, we assign the probability of creation $1 / (y_\text{u} - y_\text{l})$ to each year within this range (see figure~\ref{fig:distribs}). 

The difference in creation date of two witnesses dated in (potentially overlapping) ranges is then estimated by the expectation values of the difference between two random dates uniformly drawn from these ranges. Explicitly for two manuscripts dated in respective ranges $[a,b]$ and $[c,d]$, with $a \leq d$, the difference $D$ writes
\[ D = \frac{1}{b-a} \frac{1}{d-c} \int_a^b \int_c^d \vert x - y \vert \mathrm{d}x \mathrm{d}y\]

Summary statistics on stemmata was extracted from the DOT files available on OpenStemmata \cite{camps_open_2021b}. Beforehand, stemmata were treated to be in an homogeneous format, by removing lateral transmission, as well as superfluous nodes conventionally represented on many published stemmata (i.e. nodes standing for hypothetical lost manuscripts, but having only an out-degree of one). 

It is then possible to single out a region in the parameter space where these observable on simulated traditions coincide within an average error margin of $20\%$ to their empirical values, namely for $\lambda \in [7.5~10^{-3}, 9.5~10^{-3}]$ and $\mu \in [3.2~10^{-3}, 4.4~10^{-3}]$

\clearpage

\section*{Supplementary information}


\subsection*{Supplementary Tables}

\subsubsection*{Copy speed of scribes}

The following table contains a summary of the estimates of scribes copy speed, measured in the number of leaves (folia) per day (f/d). The estimates are gathered from the bibliography.
Publications giving estimates for a single scribe or manuscript have not been taken into consideration. A distinction has been made between professional scribes on one hand, and amateur scribes, monastic or otherwise, on the other. 
Most estimates are based on colophons, i.e., formules written by scribes at the end of the copied book, sometimes giving a date or the length of the copy. In rare cases, the information is complete (i.e., we have the start and end date, or the duration), but most often the start date has to be estimated from other evidence, such as the end date of the copy of the previous volume (in the case of multi-volume texts). On some occasions, we keep the contracts passed between the scribe and his customer, giving the number of leaves to be copied each day, or the total expected duration for the copy of a book of a given length. 
It is to be noted that most of those estimates do not substract non worked days (sundays, holidays), nor are able to account for the possibility that scribes could have worked in parallel on copying other documents, which means that the actual rate, for a full work day dedicated to a single manuscript, is expected to be higher. In addition, most of those estimates focus on the production of commercial books, and discard the speed of copy by authors of their own drafts, by students of their studying books, etc. For these productions, evidence is that the rates could be much higher (up to 10 f/d or even higher), but at the detriment of the quality of execution.
In addition, this data does not take into account variations based on the actual quantity of text per leave, dependent on the book size, number of columns and lines per columns.
The sample size N is the number of manuscripts from which the information is derived.

\begin{SupplementaryTable}[ht!]
    \centering
\caption{Estimates of manuscript copying speed}
    \begin{tabular}{ll|ccc|rr}
\textbf{Ref.} & \textbf{Source} 
& 
\textbf{Date} & \textbf{Region} &
\textbf{Type} 
& \textbf{f/d}
& \textbf{N}
\\ \hline \hline
\cite{bozzolo_pour_1980} 
& colophons 
& 1200-1600
& West. Europe 
& all
& 2.85 & 63\\ 
\cite{gumbert_speed_1995} & 
both &  medieval & West. Europe & all & 2.00 & 800\\ \hline
\cite{overgaauw_fast_1995} & colophons
& 1400-1500 & Italy
& prof.
& 2.36
& 13 \\
\cite{bozzolo_pour_1980} 
& contracts 
& 1400-1500
& Avignon 
& prof. 
& 2.74 & 3\\ \hline
\cite{overgaauw_fast_1995} & colophons
& 1400-1500
& Netherlands
& monast.
& 0.84
& 18\\ 
    \end{tabular}
    \label{tab:copy_time}
\end{SupplementaryTable}

%




\subsubsection*{Loss rates of manuscripts and incunabula}

The loss rates indicated here are drawn from the bibliography. Their original sources include, for manuscripts (\textit{MSS}), medieval catalogues (mostly of ecclesiastical institutions), where the number of original entries is compared to the number of (identifiable) surviving manuscripts; for early printed books (\textit{Pr}), the print run of early \textit{incunabula editions};
as well as, for reference, the results of different mathematical models.
The sample size (\textit{N}) is equal to the estimated total original population, before decimation.

The tables are split between a first table presenting the most relevant data to our study case (medieval vernacular chivalric literature), while the second table gives other estimates for comparison purposes.

In the second table, the final estimates of manuscript loss for the Holy Roman Empire of the German Nation and for Europe are based on the observed loss of contemporary \textit{incunabula}, while trying to accommodate for historical factors specific of manuscript production and conservation. 

\begin{SupplementaryTable}[h!]
    \centering
\caption{Estimates of loss rates of documents from various sources for chivalric vernacular texts}  \small
    \begin{tabular}{lll|cclrr|r}
\textbf{Ref} & \textbf{Type}& \textbf{Source} & 
\textbf{Date} & \textbf{Region} & \textbf{Lang.} 
& \textbf{Loss (\%)} & \textbf{N} \\ \hline \hline
\cite{camps_`chanson_2016} & MSS & catalogues & 1100-1500 & UK & Fre. & 82.00 & 45\\
\cite{trovato_everything_2014} & Pr
& print run 
& 1495-1532 & Italy 
& Ita. & 99.31 & 5200\\ \hline
\cite{kestemont_forgotten_2022} & MSS & \textit{unseen sp.} & 800-1500 & West. Eur. & vern. & $\geq$91.00 & 41\,244\\
\cite{kestemont_forgotten_2022} & MSS & \textit{unseen sp.} & 1100-1500 & West. Eur. & Fre. & $\geq$94.60 & 27\,278\\
    \end{tabular}
    \label{tab:death_rates_chivalric}
\end{SupplementaryTable}

\begin{SupplementaryTable}[h!]
    \centering
\caption{Estimates of loss rates from various sources (all other types) -- starred languages means they constitute the majority of the collection, but that some other languages may be marginally present}  \small
    \begin{tabular}{lll|cclrr|r}
\textbf{Ref} & \textbf{Type} & \textbf{Source} & 
\textbf{Date} & \textbf{Region} & \textbf{Lang.} 
& \textbf{Loss (\%)} & \textbf{N} \\ \hline \hline
\cite{buringh_medieval_2010} & MSS & catalogues & 700-1500 & UK & Lat.* & 90.83 & 17207\\
\cite{buringh_medieval_2010} & MSS & catalogues & 700-1500 & France & Lat.* & 72.01 & 11645\\
\cite{trovato_everything_2014} & Pr & print run 
& 1476-1542 & Italy 
& Ita. & 92.39 & 14700\\
\cite{trovato_everything_2014} & Pr
& print run 
& 1495-1532 & Italy & Lat. & 94.56 & 3300\\
\cite{neddermeyer_moglichkeiten_1996} & Pr & print run & 1453-1500 & Europe & all & 95.70 & 18488000\\
\cite{neddermeyer_moglichkeiten_1996} & Pr & print run & 1453-1500 & Centr. Eur. & all & 94.50 & 6144000\\ \hline
\cite{neddermeyer_moglichkeiten_1996} & MSS & \textit{estimate} & 1400-1500 &  Empire & all & $\geq$95 & 1122000\\
\cite{neddermeyer_moglichkeiten_1996} & MSS & \textit{estimate} & 1400-1500 & Europe & all & $\geq$95 & $\geq$ 2396000
    \end{tabular}
    \label{tab:death_rates_other}
\end{SupplementaryTable}

\subsubsection*{Bifidity ratio in historical stemmata collections}

The following table gives the computed bifidity ratio in stemmata collections since Bédier. When available, are additionally given the
number of trees analysed (N), the 
restrictions in terms of language(s), literary genre, status of the trees (final tree, used as basis of text editions versus provisional tree) and the criteria used for inclusion in the collection, as well as the reference to the source.

\begin{SupplementaryTable}[h!]
    \centering \small
\begin{tabular}{lllllp{3cm}l}
\textbf{bifid (\% )} & \textbf{N} & \textbf{Lang.} & \textbf{Genre restr.} & \textbf{Status} & \textbf{Collection} & \textbf{Ref.} \\ \hline \hline
95.5 & 110 & Fr, Lat, Eng, Ger & no & NA & NA & \cite{bedier_tradition_1928}\\ 
69.0 & 130 & Pro & troubadour lyric & NA & NA & \cite{shepard_recent_1930}\\
75.5 & 94 & Fr & no & all & works cited in \cite{bossuat_manuel_1951} & \cite{castellani_bedier_1957}\\ 
82.5 & 86 & Fr & no & final & works cited in \cite{bossuat_manuel_1951} & \cite{castellani_bedier_1957}\\ 
83.1 & 89 & Non & no & final  & \textit{Bibl. et Ed. Arnamagn.} & \cite{haugen_silva_2015}\\
77.0 & 117 & Fr & epics, romances & all  & \textit{OpenStemmata} & \cite{camps_open_2021b}\end{tabular}
\caption{Bifidity estimates in different stemmata collections, since Bédier}
\label{models:fig:stemmataCollections}
\label{tab:tree_props}
\end{SupplementaryTable}


\end{bibunit}

\clearpage

\section*{Declarations}

\paragraph{Acknowledgments}

The authors wish to thank Gustavo Fernandez Riva, Simon Gabay and Anna Preto for their contributions to the database of stemmata, as well as Émilie Guidi and Jade Norindr for their work on the list of witnesses. 
They additionally want to thank Nicolas Baumard, Florian Cafiero, Kelly Christensen, Katarzyna Kapitan,  Théo Moins, Olivier Morin, and Valentin Thouzeau for the feedback they provided on this research.  

\paragraph{Funding}

\noindent \includegraphics[width=0.15\textwidth, trim={1cm 7cm 1cm 0cm}, clip]{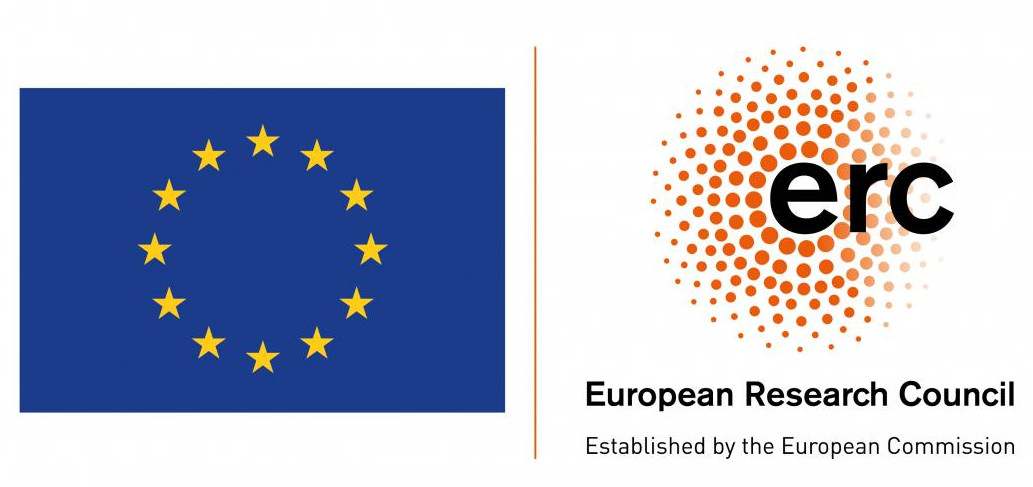}
Funded by the European Union (ERC, LostMA, 101117408). Views and opinions expressed are however those of the author(s) only and do not necessarily reflect those of the European Union or the
 European Research Council. Neither the European Union nor the granting authority can be held responsible for them.

\paragraph{Competing interests}

The authors have no competing interests to declare. 

\paragraph{Ethics approval and consent to participate}

Not applicable.

\paragraph{Consent for publication}

Not applicable.

\paragraph{Availability of data and materials}

Data and code accompanying this paper are available on Github (\url{https://github.com/LostMa-ERC/ExtinctionOfTexts})
, under a CC BY-SA 4.0 License, and will be archived on Zenodo in time of publication.

\paragraph{Code availability}

Data and code accompanying this paper are available on Github (\url{https://github.com/LostMa-ERC/ExtinctionOfTexts})
, under a CC BY-SA 4.0 License, and will be archived on Zenodo in time of publication.

\paragraph{Authors' contributions}

The authors contributed to all aspects of this publication.

\paragraph{Authors' information}

JBC is associate professor in computational philology at École nationale des chartes - PSL, as well as principal investigator of project LostMA. JRF is professor of mathematical sciences at École normale supérieure Paris-Saclay, where he coordinates the CHArt[S] (Complexity, Humanities, Art, Societies) group at the Dpt of Mathematics / Centre Borelli; he is also an affiliate professor at UM6P College of Computing and a co-lead of project LostMA. UG is post-doctoral fellow in the LostMA project.



\end{document}